\documentclass{appolb}
\usepackage{graphicx}
\usepackage{hepnames,hepunits}
\usepackage{hyperref}

\def\lesssim{\ \hbox{\raise 2pt \hbox{$<$} \kern -13pt
                     \lower 3pt \hbox{$\sim$}}\ }
\def\greatersim{\ \hbox{\raise 2pt \hbox{$>$} \kern -13pt
                     \lower 3pt \hbox{$\sim$}}\ }

\def\lsim{\mathrel{\rlap{\lower4pt\hbox{\hskip1pt$\sim$}}
    \raise1pt\hbox{$<$}}}                
\def\gsim{\mathrel{\rlap{\lower4pt\hbox{\hskip1pt$\sim$}}
    \raise1pt\hbox{$>$}}}                

\def\cascade{{\sc Cascade}}
\def\pythia{{\sc Pythia}}
\def\herwig{{\sc Herwig}}
\def\sherpa{{\sc Sherpa}}

\input epsf.tex
\def\desepsf(#1 width #2){\epsfxsize=#2 \epsfbox{#1}}
\def\kt{\ensuremath{k_{\rm T}}}

\def\PZ{\ensuremath{Z}}

\def\qt{\ensuremath{q_{\rm T}}}
\def\zdyn{\ensuremath{z_{\rm dyn}}}
\def\zm{\ensuremath{z_{\rm M}}}
\def\ptll{\ensuremath{p_{\rm T}(\ell\ell)}}

\newcommand{\sqrts}{\ensuremath{\sqrt{s}}}

\newcommand{\mdy}{\ensuremath{m_{\small\text{DY}}}\xspace}
\newcommand{\as}{\ensuremath{\alpha_s}}

\newcommand{\PBM}{PB}
\newcommand{\PBset}{{PB-NLO-2018}}

\newcommand{\MCatNLO}{{\sc MadGraph5\_aMC@NLO}}

\newcommand{\GeV}{\text{GeV}\xspace}

\newenvironment{tolerant}[1]{\par\tolerance=#1\relax}{ \par }
\usepackage{amsmath,bm}
\usepackage{lineno}

\usepackage{cite}

\begin{document}
\title{The non-perturbative Sudakov Form Factor and the role of soft gluons.%
\thanks{Presented at Cracow Epiphany Conference 2024}%
}
\author{Hannes Jung
\address{Deutsches Elektronen-Synchrotron DESY, Germany, \\
II. Institut f\"ur Theoretische Physik, Universit\"at Hamburg,  Hamburg, Germany}
}
\maketitle
\begin{abstract}
The role of soft gluons in inclusive collinear parton densities as well as in Transverse Momentum Dependent (TMD) parton densities is discussed. Applying  the Parton-Branching (\PBM ) method,  the so-called non-perturbative Sudakov form factor could be identified with the integration range $z \to 1$, which is neglected in collinear parton shower approaches.

The importance of soft gluons could be shown by investigating the transverse momentum spectrum of Drell-Yan lepton pairs, leading to a width of the intrinsic-\kt\ distribution which is independent on \sqrts , in contrast to what is observed in parton shower approaches. The reason for this behavior is traced back to the non-perturbative Sudakov form factor. The role of soft gluons for observable hadron spectra is discussed and shown to be negligible.

This talk is dedicated to the memory of S. Jadach.

\end{abstract}
  
\section{Introduction}
The transverse momentum of Drell-Yan (DY) lepton pairs \ptll\ has been measured with great precision over a wide range in  \sqrts\  (from 30~\GeV\  to LHC energies e.g.~\cite{Ito:1980ev,Moreno:1990sf,Aidala:2018ajl,Aad:2015auj,CMS:2022ubq}). At large \ptll\ the spectrum is described by next-to-leading (NLO) perturbative calculations, at smaller \ptll\ soft gluon radiation has to be resummed to all orders, and at very small \ptll\ non-perturbative contributions play a role. The resummation is performed in the form of CSS~\cite{Collins:1984kg} or transverse momentum dependent (TMD) factorization~\cite{Collins:2012ss} or in form of parton showers in multipurpose Monte Carlo event generators like \herwig~\cite{Bellm:2015jjp,Bahr:2008pv},  \pythia~\cite{Sjostrand:2014zea,Bierlich:2022pfr} or \sherpa~\cite{Bothmann:2019yzt,Gleisberg:2008ta}. 

Especially the region of very small \ptll\ is of great interest, since this region affects very much the determination of the \PW -mass via the transverse momentum spectrum of the decay lepton.
In CSS and TMD factorization the small \ptll -region is described by  a contribution of non-perturbative intrinsic transverse momentum of partons inside the hadron and a so-called non-perturbative Sudakov form factor~\cite{Hautmann:2020cyp,Wei:2020glg,Scimemi:2019cmh}. In Monte Carlo event generators the parton shower treats parton emissions which are resolvable (e.g. with a transverse momentum larger than a cutoff, $q_0$, of the order of \GeV ) in addition to intrinsic transverse momentum distributions. However, in order to describe the low \ptll\  region of the spectrum at different \sqrts , the width of the intrinsic transverse momentum distribution, parametrized as a Gauss distribution, is required to be \sqrts\ dependent~\cite{Sara-PIC2023}. On the contrary, the Parton Branching (\PBM ) approach~\cite{Hautmann:2017fcj,Hautmann:2017xtx} with its implementation into the Monte Carlo event generator \cascade 3~\cite{Baranov:2021uol} allows a description of the \ptll\ spectrum at small \ptll\ with a width of the intrinsic-\kt\ distribution which is independent on \sqrts ~\cite{Bubanja:2023nrd}

The \PBM~method provides an intuitive and direct bridge between CSS and parton shower approaches in Monte Carlo event generators and therefore offers the direct possibility to study their relation in detail and especially to understand the origin of the so-called non-perturbative Sudakov form factor~\cite{Mendizabal:2023mel}.

\section{Parton Branching method and non-perturbative Sudakov form factor}
The Parton Branching method~\cite{Hautmann:2017fcj,Hautmann:2017xtx}  was developed as a solution to the DGLAP evolution equations and to provide information on the transverse momentum distributions of the evolving partons by applying a reformulation of the DGLAP evolution equation in terms of Sudakov form factors (see e.g. Ref.~\cite{Ellis:1991qj}). A similar method has been developed in Cracow group around S. Jadach \cite{GolecBiernat:2007pu,GolecBiernat:2006xw,Jadach:2008nu,Jadach:2003bu,Placzek:2007xb}.

The Sudakov form factor $\Delta_a ( \mu^2 , \mu^2_0 )$ is essential in the formulation of the \PBM -method:
\begin{equation}
\label{sud-def}
 \Delta_a ( \mu^2 , \mu^2_0 ) = 
\exp \left(  -  \sum_b  
\int^{\mu^2}_{\mu^2_0} 
{{d {\bf q}^{\prime 2} } 
\over {\bf q}^{\prime 2} } 
 \int_0^{\zm} dz \  z 
\ P_{ba}^{(R)}\left(\alpha_s , 
 z \right) 
\right) 
  \;\; ,   
\end{equation}
with resolvable splitting functions $P_{ba}^{(R)}\left(\alpha_s ,z \right)$  for splitting of parton $a$ into parton $b$ as a function of  the splitting variable $z$ which is the ratio of the longitudinal momenta of the involved partons. The splitting functions are the DGLAP splitting functions at leading or next-to-leading order. For numerical stability, the parameter \zm\ is introduced  with $\zm = 1 - \epsilon$ with $\epsilon \to 0$. It is essential that $\epsilon \to 0$ to ensure proper cancellation of the different terms in the derivation of the evolution equation as well as to reproduce the DGLAP limit as shown in Ref.~\cite{Hautmann:2017fcj,Hautmann:2017xtx}  and to obtain stable solutions for TMD distributions. 

The \PBM\ evolution equation for a TMD density $ { {\cal A}}_a(x,{\bf k}, \mu^2)$ for parton $a$ at scale $\mu$  can be written in integral form as:
\begin{eqnarray}
\label{evoleqforA}
   { {\cal A}}_a(x,{\bf k}, \mu^2) 
 &=&  
 \Delta_a (  \mu^2  ) \ 
 { {\cal A}}_a(x,{\bf k},\mu^2_0)  
 + \sum_b 
\int_{\mu^2_0}^{\mu^2}
{{d^2 {\bf q}^{\prime } } 
\over {\pi {\bf q}^{\prime 2} } }
 \ 
{
{\Delta_a (  \mu^2  )} 
 \over 
{\Delta_a (  {\bf q}^{\prime 2}  
 ) }
}
 \nonumber\\ 
 & &\times  
\int_x^{\zm } {{dz}\over z} \;
P_{ab}^{(R)} (\alpha_s 
,z) 
\;{ {\cal A}}_b\left({x \over z}, {\bf k}+(1-z) {\bf q}^\prime , 
{\bf q}^{\prime 2}\right)  
  \;\;  ,     
\end{eqnarray}
with $x$ being the longitudinal momentum fraction and ${\bf k}$ being the 2-dimensional vector of the transverse momentum with $\kt = |{\bf k}|$ and $| {\bf q}^\prime| =q'$.

The starting distribution ${ {\cal A}}_a(x,{\bf k},\mu^2_0)$ in eq.(\ref{evoleqforA}) at scale $\mu_0$ is parametrized in terms of a  collinear parton density at the starting scale and the intrinsic-\kt\ distribution described as a  Gaussian distribution of  width $\sigma$:
\begin{equation}
\label{TMD_A0}
{\cal A}_{0,a} (x, {\bf k},\mu_0^2)   =  f_{0,a} (x,\mu_0^2)  
\cdot \exp\left(-\kt^2 / 2 \sigma^2\right) / ( 2 \pi \sigma^2) \; .
\end{equation}
The width of the Gaussian distribution $\sigma$ is related to the parameter $q_s$ defined by  $q_s = \sqrt{2} \sigma$. 

For the evolution of the transverse momentum, angular ordering is essential, which relates  the evolution scale  $q'$  to the transverse momentum \qt\ of the emitted parton via : 
\begin{equation}
q' = \qt / (1-z) \; .
\label{angord}
\end{equation}

\begin{tolerant}{8000}
The \PBM\ evolution equation has been used to determine collinear and TMD distributions by fits to deep-inelastic measurements at HERA~\cite{Martinez:2018jxt}. One parametrization (\PBset~Set1) reproduced the HERAPDF collinear distribution with the evolution scale $q'$ as scale in \as . Angular ordering suggests that \qt\ should be used as a scale in \as , which has been done for \PBset~Set2. 
This choice required to define two different regions: a perturbative region, with $\qt > q_0$, and a non-perturbative region of $\qt < q_0$, where \as\ is frozen at $q_0$.
\end{tolerant}

The requirement of the perturbative region, $\qt > q_0$, leads directly to a restriction of $z$ as given by eq.(\ref{angord}):
\begin{equation}
  \label{zdyn} 
\zdyn = 1 - q_0/q' \; .
\end{equation}

The Sudakov form factor can now be separated into a perturbative ($0 < z < \zdyn$) and non-perturbative ($\zdyn < z < \zm$) part~\cite{Mendizabal:2023mel,Martinez:2024twn}:
\begin{eqnarray}
\label{eq:divided_sud}
 \Delta_a ( \mu^2 , \mu^2_0 ) & = &
\exp \left(  -  \sum_b  
\int^{\mu^2}_{\mu^2_0} 
{{d {\bf q}^{\prime 2} } 
\over {\bf q}^{\prime 2} } 
 \int_0^{\zdyn} dz \  z 
\ P_{ba}^{(R)}\left(\alpha_s , 
 z \right) 
\right) \nonumber \\
& & 
 \times \exp \left(  -  \sum_b  
\int^{\mu^2}_{\mu^2_0} 
{{d {\bf q}^{\prime 2} } 
\over {\bf q}^{\prime 2} } 
 \int_{\zdyn}^{\zm} dz \  z 
\ P_{ba}^{(R)}\left(\alpha_s , 
 z \right) 
\right) \nonumber \\
& = &  \Delta_a^{(\text{P})}\left(\mu^2,\mu_0^2,q^2_0\right)  \cdot \Delta_a^{(\text{NP})}\left(\mu^2,\mu_0^2,q_0^2\right) \; .
\end{eqnarray}

\begin{figure}[htb]
\centerline{
\includegraphics[width=0.49\textwidth]{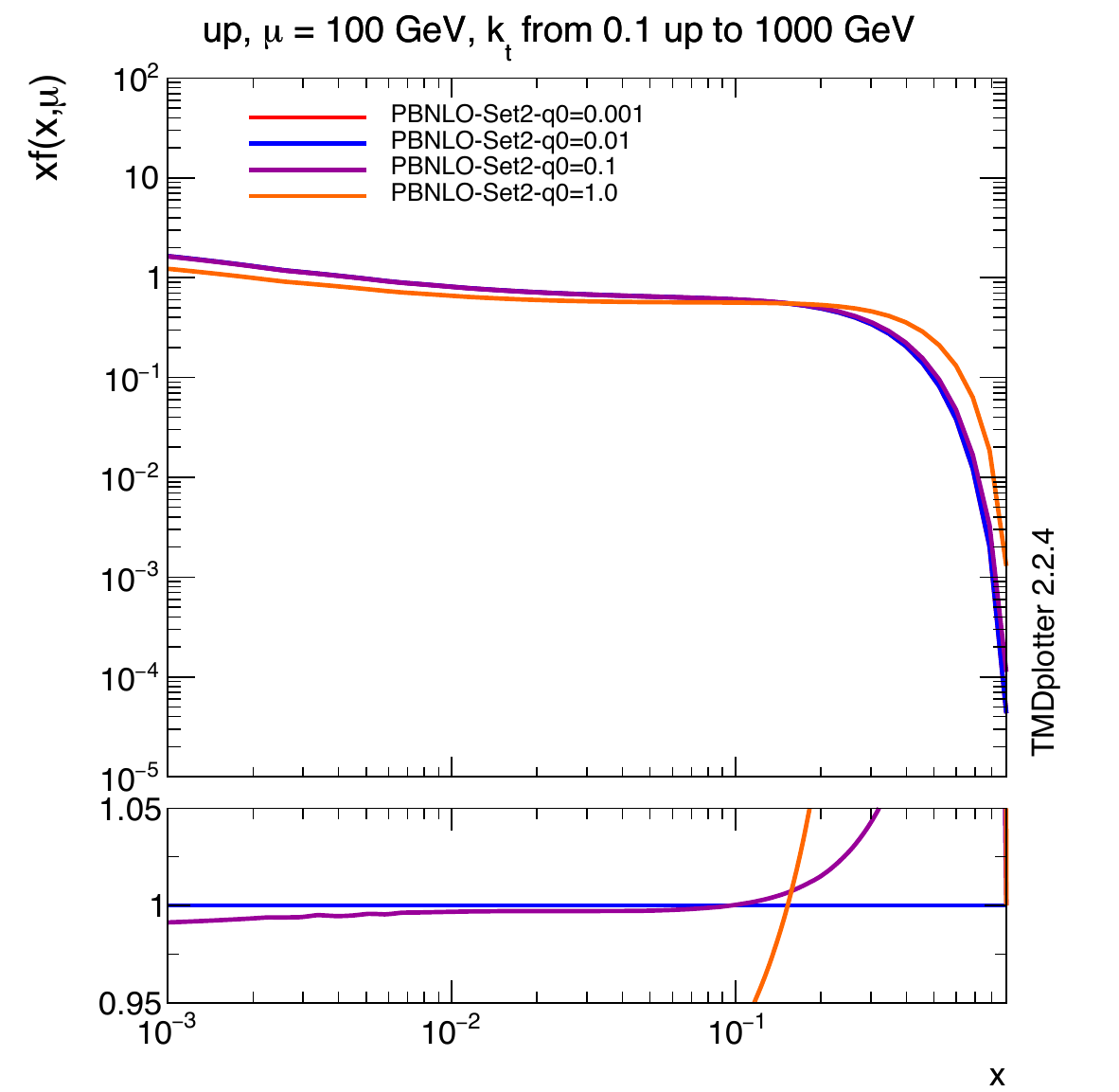}
\includegraphics[width=0.49\textwidth]{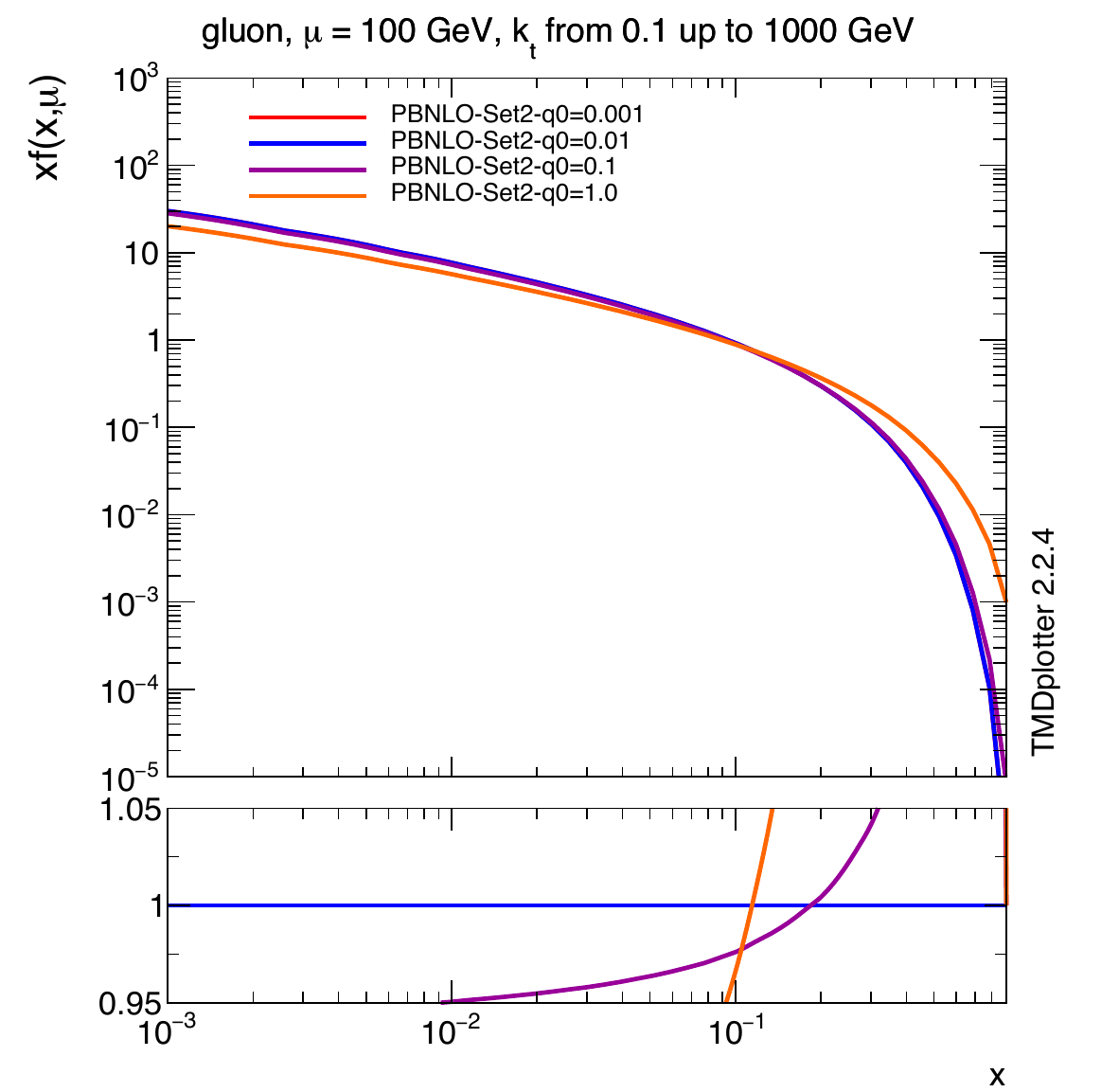}
}
\caption{Parton density of up-quark and gluon at $\mu=100$~\GeV\ for different values of~$q_0$ (used in $\zdyn = 1 - q_0/q' $) obtained with the same initial parameters as for \PBset~Set2. }
\label{zm_limit}
\end{figure}

In Fig.~\ref{zm_limit} integrated quark and gluon distributions at a scale of $\mu=100$~\GeV\ are shown for different values of $q_0$ which limits the $z$-integration with $\zdyn = 1 - q_0/q' $. 
Stable results are obtained for $q_0 < 0.01$ \GeV\  while differences are visible in inclusive distributions for $q_0 > 0.01$ \GeV, showing the importance of the non-perturbative region even for inclusive distributions.

The \PBM -method allows to make a connection to the CSS formalism: applying angular ordering  with  $\zdyn = 1 - q_0/q' $  together with the transverse momentum $\qt = q' (1-z)$ taken as a scale in \as\ allows to obtain the different terms of the CSS Sudakov, even up to next-to-next-to leading order as discussed in Ref.~\cite{Martinez:2024twn}.  
In addition, the great advantage of the \PBM -method is that the  non-perturbative Sudakov form factor $\Delta_a^{(\text{NP})}$  can be calculated and is fixed by the fit to inclusive distributions~\cite{Martinez:2018jxt}, in contrast to CSS, where this form factor has to be constrained  additionaly from exclusive measurements.

\subsection{Soft gluons and the  \ptll\ spectrum in Drell-Yan production}
The transverse momentum \ptll\  spectrum of DY lepton pairs is the benchmark measurement for TMD pdfs, soft-gluon  resummation and parton shower approaches. While the spectrum at large \ptll\ is well described by perturbative calculations at NLO, at smaller \ptll\ TMDs, soft gluon resummation or initial state parton showers are needed, at very small \ptll\ non-perturbative effects play a role. In collinear parton shower approaches, this very low \ptll\ region is described by the inclusion of  intrinsic motion of partons inside the hadrons, usually described by a Gauss distribution. In the shower description of \pythia\ or \herwig\ a minimum transverse momentum of the radiated parton is required, leading essentially to a suppression of soft gluon emissions in the non-perturbative region. 
In the \PBM -approach, the soft gluon region is covered by the requirement that $\zm = 1 - \epsilon$ with $\epsilon \to 0$.

\begin{figure}[htb]
\centerline{
\includegraphics[width=0.5\textwidth]{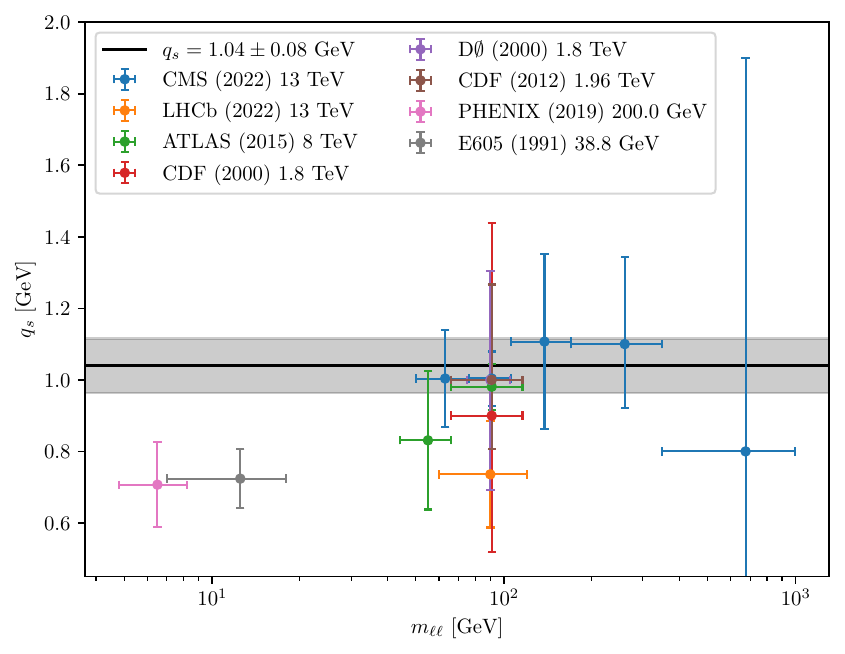}
\includegraphics[width=0.5\textwidth]{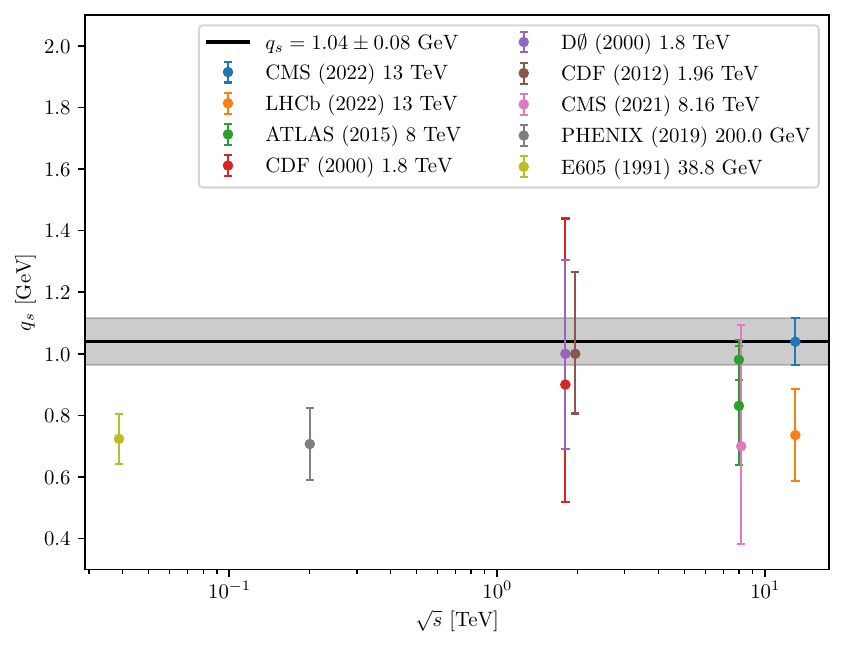}
}
\caption{The width parameter $q_s$ of the intrinsic-\kt\ distribution as a function of the DY mass \mdy\ and  the center-of-mass energy \sqrts . (Plots taken from Ref.~\cite{Bubanja:2023nrd}) }
\label{PB-kt}
\end{figure}

In Ref.~\cite{Bubanja:2023nrd}, the transverse momentum spectrum \ptll\ is studied and the width of intrinsic-\kt\ distribution is determined from precision measurements at LHC energies~\cite{CMS:2022ubq} over a wide range of DY masses \mdy . Measurements at lower energies were also investigated, and it was found, that all measurements can be reasonably well described with the \PBM -approach applying \PBset~Set2 with a width of the intrinsic-\kt\ distribution which is independent of the DY mass \mdy\ as well as of \sqrts\ as shown in Fig.~\ref{PB-kt}.

\begin{figure}[htb]
\centerline{
\includegraphics[width=0.7\textwidth]{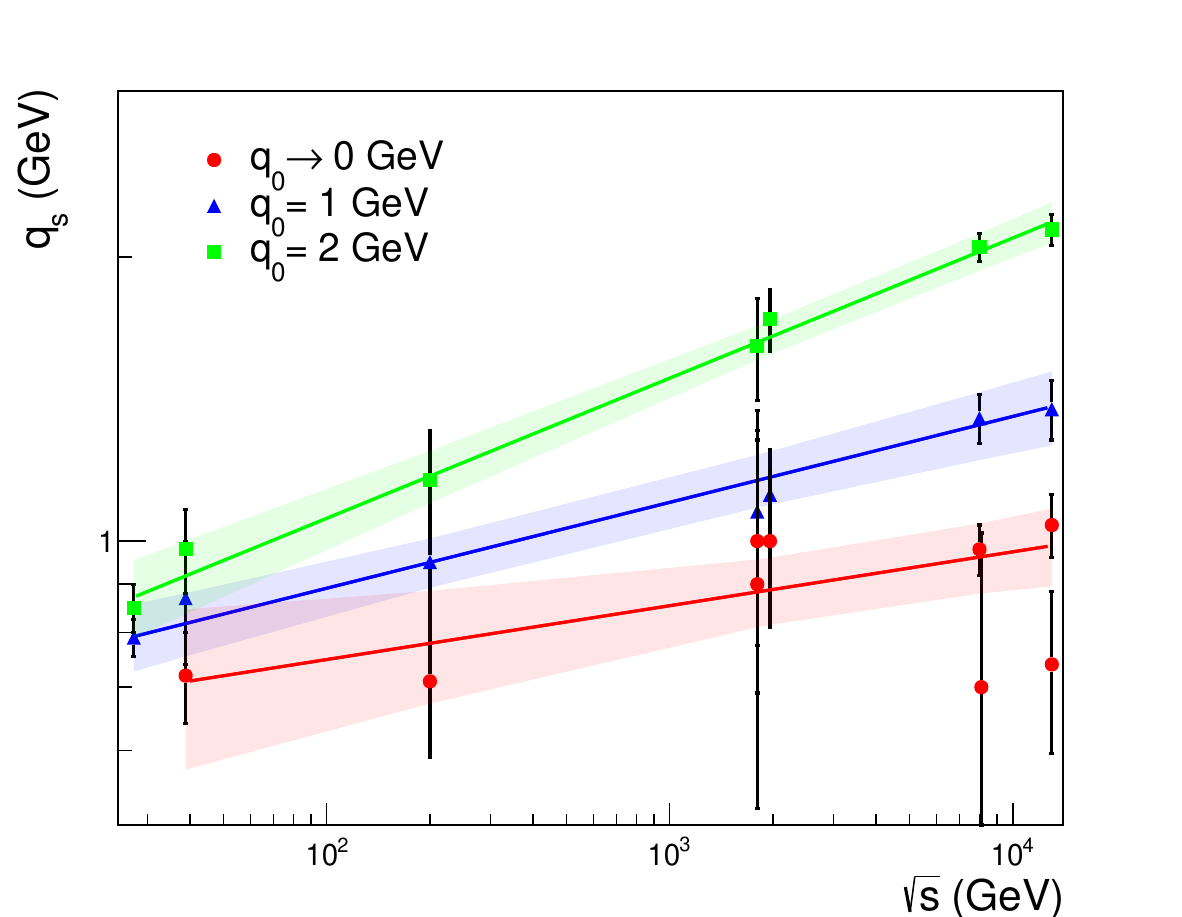}
}
\caption{The width parameter $q_s$ of the intrinsic-\kt\ distribution as a function of center-of-mass energy \sqrts . The uncertainty bands show the 95\% CL. 
 (Plots taken from Ref.~\cite{Bubanja:2024puv}) }
\label{kt-width}
\end{figure}

The width of the intrinsic-\kt\ distribution to be used in the collinear parton shower Monte Carlo event generators \pythia\ and \herwig\ has been determined in Ref.~\cite{CMS-PAS-GEN-22-001}, and a strong dependence on the center-of-mass energy \sqrts\ has been observed, in contrast to the findings in Ref.~\cite{Bubanja:2023nrd}. In order to understand this behavior, a dedicated study has been performed in Ref.~\cite{Bubanja:2024puv}. The \PBM -method has been used to mimic the behavior of collinear parton shower Monte Carlo event generators by imposing a restriction on $z$ via $\zdyn = 1 - q_0/q' $:  in \herwig\  angular ordering is applied (similar to \PBM ) and the parameter 
$Q_{g,q}$~\cite{Bahr:2008pv}[p~659] restricts the $z$-integration range and  in \pythia\   $z_{max}(Q^2)$ \cite{Bierlich:2022pfr}[p~60] is used.
These restrictions remove completely  $\Delta_a^{(\text{NP})}$ from eq.(\ref{eq:divided_sud}).  Calculations have been performed using \MCatNLO~\cite{Alwall:2014hca} together with \PBM\ collinear parton distributions of \PBset~Set2. The \PBM -TMD distributions have been recalculated imposing two different cuts on $z$ by  $\zdyn = 1 - q_0/q' $ with $q_0=1(2)$\GeV . The collinear distributions were left unchanged in order to keep consistency with the NLO hard process calculation, as argued in Ref.~\cite{Mendizabal:2023mel}. The width of the intrinsic-\kt\ distribution has been determined from different DY measurements over a large range of center-of-mass energy \sqrts .
The obtained widths of the intrinsic-\kt\ distributions were found to be strongly dependent on the center-of-mass energy, as shown in Fig.~\ref{kt-width}: a strong dependence of the width parameter $q_s$ on the center of mass energy \sqrts\ is observed as soon as the $z$ integration range is restricted by $\zdyn = 1 - q_0/q' $ with a stronger dependence being observed for larger values of $q_s$. For comparison also the results obtained in Ref.~\cite{Bubanja:2023nrd} are shown. The large uncertainty of the straight line fit indicates, that even a constant line is acceptable for $q_s \to 0$, while for $q_s=2$~\GeV\ a steep slope is visible.

The results of Ref.~\cite{Bubanja:2024puv,Mendizabal:2023mel,Bubanja:2023nrd} prove that soft, non-perturbative gluons play an important role for a  consistent description of inclusive distributions, such as integrated collinear parton densities as well as for 
the transverse momentum distribution \ptll\ of DY lepton pairs.

\subsection{Soft gluons and hadronic final states}

Non-perturbaitve, soft gluons, with transverse momenta below a \GeV\ (and even below $\Lambda_{QCD}$), are clearly not detectable and measurable as partons or jets. Such soft contributions play an important role in inclusive distributions, as shown in the previous sections. The effect of these soft gluons has been studied Ref.~\cite{Mendizabal:2023mel} within the \PBM -TMD parton shower simulation in \cascade3~\cite{Baranov:2021uol}. As for inclusive distributions sensitive to the TMD distributions, the \PBM -TMD parton shower can be used to simulate the effects of soft gluons by applying different cuts on $q_0$ in $\zdyn = 1 - q_0/q' $. 

\begin{figure}[htb]
\centerline{
\includegraphics[width=0.5\textwidth]{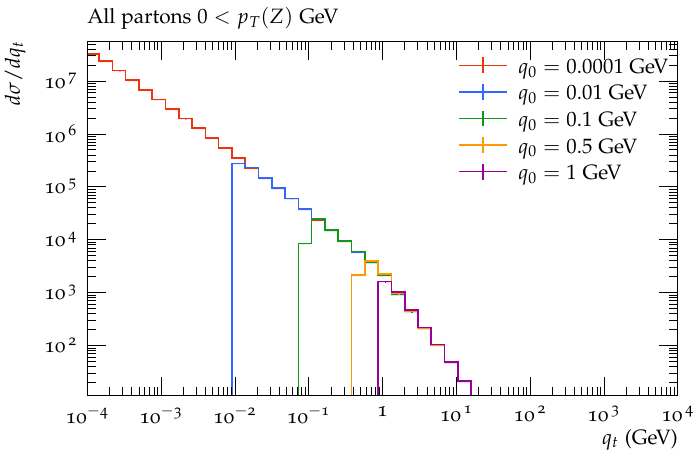}
\includegraphics[width=0.5\textwidth]{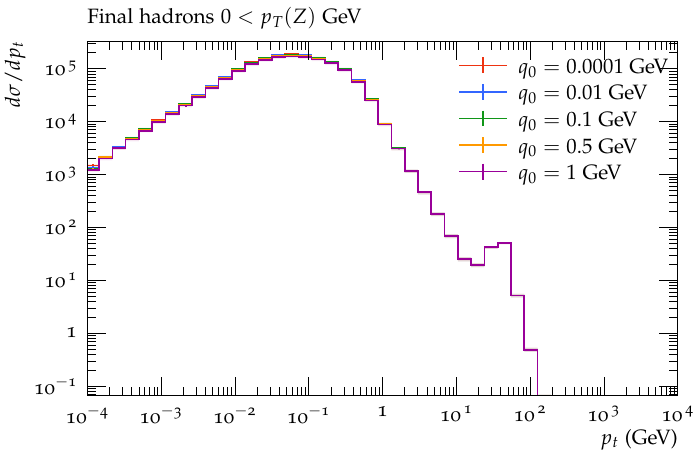}
}
\caption{Distributions of the transverse momentum distributions of emitted partons in the initial state cascade (left) and observable hadrons (right)  for different values of $q_0$ in \PZ -boson events. 
 (Plots taken from Ref.~\cite{Mendizabal:2023mel}) }
\label{qt-pthad}
\end{figure}

While clearly the accessible  $z$-range is affected by different choices of $q_0$ and obviously the range of  transverse momenta \qt\ of emitted partons during the initial state cascades (shown in Fig.~\ref{qt-pthad}(left)), the spectrum of observable hadrons is not affected at all (Fig.~\ref{qt-pthad}(right)). In the Lund string fragmentation, gluons act as kinks in the color string, and therefore these soft gluons play a negligible role. In  cluster fragmentation, as used in \herwig\ a finite transverse momentum is required to form colorless clusters, and soft gluons cannot be included easily.

\section{Conclusion}
The \PBM -approach offers a very intuitive and direct link between analytic resummation for inclusive parton distributions obtained with the DGLAP equation on one hand and  TMD resummation  and parton shower approaches on the other hand. 
It was shown that on an inclusive level, the \PBM -approach reproduces exactly semi-analytical solutions of  the DGLAP evolution equation, as long as the $z$-integration is covering the range of soft gluons with $z \to 1$. This requirement is also essential to ensure proper cancellation of different terms when cross sections are calculated at NLO. 

The \PBM -method provides directly also TMD distributions, since every single branching is simulated and kinematic relations can be included. A transverse momentum can be defined with a physical interpretation of the evolution scale, in the \PBM -method angular ordering is applied. From the \PBM -Sudakov form factor the different terms of the CSS formalism can be directly obtained (even at NNLO) if the $z$-integration range is restricted to perturbative gluons via $\zdyn = 1 - q_0/q' $. 
A feature of the \PBM -Sudakov form factor is, that it covers also the region of $z > \zdyn$ and the so-called non-perturbative Sudakov form factor, which is introduce by hand in CSS, is obtained with its free parameters fixed from the fit to inclusive collinear distributions.

Non-perturbative, soft gluons are essential for a description of the transverse momentum \ptll\ spectrum of  Drell-Yan lepton pairs at very low \ptll : only by including the non-perturbative Sudakov form factor a width of the intrinsic-\kt\ distribution  is obtained which is \sqrts\ independent, in contrast to what is observed in collinear parton shower simulations. The \PBM -method has been used to explain this behavior as coming from the restriction to perturbative gluon emission in  parton shower approaches.
With a simulation of \PBM -parton showers and the Lund String fragmentation it could be shown, that the soft gluons only matter for inclusive distributions, while there is no effect on observable hadron spectra.

\section*{Acknowledgements}
I am grateful to many inspiring, motivating and exciting discussions with S. Jadach on the Monte Carlo solution of evolution equations and transverse momentum dependent distributions. His intuition and expertise as well as his scientific achievements were always fascinating for me.

I am also grateful to my colleagues from the \cascade\ group and in particular to Itana Bubanja, Aleksandra Lelek, Mikel Mendizabal, Natasa Raicevic and Sara Taheri Monfared.
I also thank the Epiphany 2024 organizing committee for providing support to attend this conference.

\bibliographystyle{mybibstyle-new.bst}
\raggedright  
\providecommand{\href}[2]{#2}\begingroup\raggedright\endgroup

\end{document}